\documentclass{iopconfser}

\usepackage{graphicx}
\usepackage{dcolumn}
\usepackage{bm}
\usepackage{amsfonts}
\usepackage{amsmath}
\usepackage{enumitem}
\usepackage[merge,numbers,sort,compress]{natbib}

\newcommand{\Q}{\mathbb{Q}}

\newcommand{\F}{\mathbb{F}}
\newcommand{\bigO}{\mathcal{O}}

\begin{document}

\title{Rational-function interpolation from $p$-adic evaluations in scattering amplitude calculations}

\author{Herschel A. Chawdhry}

\affil{Department of Physics, Florida State University, 77 Chieftan Way, Tallahassee FL, USA}

\email{hchawdhry@fsu.edu}

\begin{abstract}
Numerical interpolation techniques are widely employed for calculating large rational functions in scattering amplitude computations.
It has been observed in recent years that these rational functions greatly simplify upon partial fractioning.
In this conference proceedings paper, based on the article~\cite{Chawdhry:2023yyx}, a technique is presented to interpolate such rational functions directly in partial-fractioned form, from evaluations at special integer points chosen for their properties under a $p$-adic absolute value.
It is shown that the technique can require 25 times fewer numerical probes than conventional finite-field-based techniques and can produce results that are more compact in size by 2 orders of magnitude.
The reconstructed results moreover exhibit additional patterns that could be exploited in future work to further improve the size of the results and the number of required numerical probes.

\end{abstract}

\section{Introduction}\label{sec:intro}
Multi-loop scattering amplitudes are core ingredients in high-precision predictions for particle collider processes. When calculating these amplitudes, a key computational bottleneck is the evaluation of elementary arithmetical operations acting on rational functions.

While performing elementary arithmetic on rational functions is conceptually straight-forward, in practice the rational functions become large at intermediate stages of calculations (compared to intial and final stages) and so the arithmetic becomes very slow.
This phenomenon is ubiquitous in computer algebra and in that field it has been common since the 1960s to instead perform these arithmetical operations numerically at sample points in a finite field $\F_p$ or a $p$-adic field $\Q_p$ and then obtain the final result in full analytic form by interpolating from several numerical samples. The computational cost of such an approach is determined primarily by the cost of obtaining sufficiently many numerical samples. Since only the final result is interpolated, the number of required numerical samples depends only on the complexity of the final symbolic result and hence bypasses the manipulation of large intermediate expressions. The use of finite fields $\F_p$ (i.e. integer arithmetic performed modulo a prime number $p$) instead of floating-point real numbers ensures that floating-point rounding errors are not introduced, and so the final interpolated answer is exactly correct.
In the last few years, finite-field methods have been adopted in the scattering amplitudes community with much success~\cite{vonManteuffel:2014ixa,Peraro:2016wsq,Klappert:2019emp,Peraro:2019svx,Klappert:2020aqs}.

The rational functions interpolated using finite-field methods are typically obtained in common-denominator form, i.e. as a ratio of two polynomials.
In multi-loop calculations in recent years, it has been observed that these rational functions, once interpolated, can be simplified by up to 2 orders of magnitude by partial fractioning them~\cite{Abreu:2019odu,Boehm:2020ijp,Agarwal:2021grm,Agarwal:2021vdh,Bendle:2021ueg,Heller:2021qkz,Chawdhry:2021mkw,Badger:2022mrb}.
In principle, it would be desirable and advantageous to exploit this simplification earlier, i.e. \emph{during} the reconstruction, so as to reduce the number of probes required by up to 2 orders of magnitude.
But as we show in more detail in our main article~\cite{Chawdhry:2023yyx}, the simplification under partial fractioning seems not to be a generic feature of rational functions, but instead a special property of the specific rational functions that appear in scattering amplitude calculations.
To exploit this simplification therefore requires developing specialised techniques that go beyond those used in generic computer algebra calculations.
This is in contrast to the use of finite-field methods, which are the computer algebraist's standard solution to the widespread computer algebra phenomenon of intermediate-expression swell.

In a scattering-amplitudes context, some work has previously been performed with the aim of optimising numerical reconstruction methods.
It is useful to guess~\cite{Abreu:2018zmy,Heller:2021qkz} the common denominator of the rational functions in multi-loop calculations, thereby reducing the number of required numerical probes by a factor of 2.
Refs~\cite{Laurentis:2019bjh,DeLaurentis:2020qle} interpolate in partial-fractioned form using very high-precision floating-point evaluations.
Within a finite-field context, it can be beneficial to reconstruct in one variable at a time and perform single-variable partial fractioning at some intermediate stages~\cite{10.1145/800206.806398,Bendle:2019csk,Badger:2021imn,Badger:2021nhg,Badger:2021ega,Abreu:2021asb}, possibly in conjunction with expansions in $\epsilon$, where $D=4-2\epsilon$ is the spacetime dimension variable.
Techniques based on algebraic geometry and evaluations in $\Q_p$ have been proposed~\cite{DeLaurentis:2022otd,DeLaurentis:2023nss,DeLaurentis:2023izi} for eliciting information about the numerator of a rational function prior to performing a finite-field reconstruction, and Ref.~\cite{Campbell:2022qpq} mentions combining these with the methods of Ref.~\cite{Laurentis:2019bjh}.

In this brief conference proceedings paper, based on our article~\cite{Chawdhry:2023yyx} to which we refer the interested reader for more comprehensive explanations, we discuss a new technique to reconstruct rational functions directly in partial-fractioned form.
Our technique uses $p$-adic probes to reconstruct the rational functions one partial-fractioned term at a time, exploiting the simplification under partial fractioning and exposing hints of further patterns and structure.
We demonstrate our technique by applying it to interpolate the largest rational function appearing in the largest IBP~\cite{Tkachov:1981wb,Chetyrkin:1981qh} expression used in Ref.~\cite{Agarwal:2021vdh} for calculating the complete set of full-colour 2-loop amplitudes for $pp \rightarrow \gamma \gamma j$ in massless Quantum Chromodynamics (QCD).
We will denote this function $R_*$.
This is a highly complicated example which is at the edge of the capabilities of current multi-loop tools and methods.


\section{Method}\label{sec:method}
In this proceedings paper we will summarise a few features of our reconstruction method before proceeding to discuss results. A more comprehensive description of our method can be found in our main article~\cite{Chawdhry:2023yyx}.

Our aim is to reconstruct a rational function $R$ in partial-fractioned form
\begin{equation}
R = \sum_i \frac{n_i}{d_i}.
\end{equation}
For reasons explained in our main article~\cite{Chawdhry:2023yyx}, it is straight-forward to generate a list of candidate denominators $\{d_i\}$ that could appear in the partial-fractioned form any particular rational function $R$.
The computational cost of interpolating in partial-fractioned form is therefore determined by the cost of obtaining sufficiently many numerical samples to interpolate all of the numerators $\{n_i\}$.
Much of the simplification upon partial fractioning seems to arise because for many of these $d_i$, the corresponding numerator $n_i$ is zero.
In other words, many candidate partial-fractioned terms vanish.
Curiously, this vanishing appears to be a particular property of the rational functions that appear in loop calculations.
Further details on these statements can be found in our main article~\cite{Chawdhry:2023yyx}.

Our method therefore aims to perform the interpolation one partial-fractioned term $\frac{n_i}{d_i}$ at a time. This ensures that if a particular partial-fractioned term vanishes, it can be cheaply detected and so we can avoid the computational cost of interpolating its numerator $n_i$. Performing the reconstruction one partial-fractioned term at a time has many other advantages that are discussed in our main article~\cite{Chawdhry:2023yyx}.

The numerical evaluations will be performed using $p$-adic numbers because as highlighted in Ref.~\cite{DeLaurentis:2022otd} they are well suited to studying the singular limits of polynomials and rational functions, which is helpful for studying the partial-fractioning of rational functions as desired in this work. Our article~\cite{Chawdhry:2023yyx} provides a brief introduction to the $p$-adic numbers and summarises the properties on which our method depends.

Let $N$ denote the number of variables that $R$ depends upon. The starting point of our method is the observation that if we can find a $p$-adic point $\overline{x} \in \Q_p ^N$ that makes one of the candidate denominators $d_k$ become $p$-adically smaller than all of the other candidate denominators $\{d_i\ : i\neq k\}$, then evaluating $R$ at that point will give us a $p$-adic series
\begin{equation}
R(x) = \frac{n_k(\overline{x})}{d_k(\overline{x})} + \bigO\left(\frac{1}{p^{m-1}}\right).
\end{equation}
Thus by considering the $p$-adic expansion of $R(\overline{x})$ we can see  whether $n_k$ is one of the candidate numerators that vanishes, in which case we can avoid interpolating it.
If $n_k$ is non-vanishing, we can perform a few more samples, always choosing points that pick out the same denominator $d_k$, and then interpolate the polynomial $n_k$.
Initially, the coefficients in $n_k$ will be obtained modulo a particular prime number $p$ corresponding to the $p$-adic field used for the samples.
One must then repeat using other values of $p$ and use the Chinese remainder theorem to obtain complete expression for $n_k$.
Note that it is important to do this before proceeding to consider other candidate partial-fractioned terms.

There are many details that have been omitted from this brief conference proceedings paper but are explained comprehensively in our main article~\cite{Chawdhry:2023yyx}. This includes, for example, explanations of how to generate points that pick out only one candidate denominator $d_k$. It also includes discussions of some possible methods for implementing $p$-adic evaluations on a computer and their respective costs relative to conventional finite-field evaluations.

\section{Results}\label{sec:results}

To demonstrate our $p$-adic interpolation technique, we employed it to calculate the large rational function $R_*$ (introduced in sec.~\ref{sec:intro}) from numerical evaluations, using no prior knowledge about it other than its mass dimension and its (easily-obtainable) common denominator.
Our interpolation technique makes no approximations and we confirmed that the interpolated result, although different in size, is mathematically exactly equal to the original expresssion.
Table~\ref{tab:results} shows that the reconstructed result is 134 times smaller than the common-denominator form that conventional finite-field methods would produce.
\begin{table}
\caption{
\label{tab:results}
Comparison of original and reconstructed form of $R_*$.
Original expression is in common-denominator form, with numerator fully expanded and denominator fully factorised.
Sizes are as reported using \texttt{ByteCount} in \textsc{Mathematica}.
Number of free parameters is obtained by counting the number of terms in the fully-expanded numerator(s). 
}
\begin{center}
\begin{tabular}{c|c|c}
Expression	&	Size	& Parameters to fit \\
\hline
Original & 605 MB &	1,369,559 \\
Reconstructed	& 4.5 MB	& 52,527 (\emph{of which 15,403 non-zero})
\end{tabular}
\end{center}
\end{table}
We required around $6*10^4$ $p$-adic probes (per prime) whereas conventional finite-field methods would require $1.4*10^6$ finite-field probes (per prime).
The free parameters fitted in the partial-fractioned expression have relatively simple values, and for this reason for most of the partial-fractioned terms we only required 3 or 4 primes, plus one more for checks, although occasionally we required as many as 10 primes.

There are interesting patterns and features in the reconstructed result, and in future work it may be beneficial to study and exploit them.
These patterns can be seen by considering, as an example, the following partial-fractioned terms
\begin{multline}\label{eq:some_reconstructed_terms}
\frac{\frac{45}{1024} s_{45}^6 s_{12}^3}{(D-3) s_{34}^4 s_{51} (-s_{23}+s_{45}+s_{51})^3}
+\frac{\frac{9}{5120} s_{45}^6 s_{12}^3}{(D-1) s_{34}^4 s_{51} (-s_{23}+s_{45}+s_{51})^3}\\
-\frac{\frac{693}{5120} s_{45}^6 s_{12}^3}{(2 D-7) s_{34}^4 s_{51} (-s_{23}+s_{45}+s_{51})^3}
-\frac{\frac{3}{1024} s_{45}^6 s_{12}^3}{s_{34}^4 s_{51} (-s_{23}+s_{45}+s_{51})^3}\\
+\frac{-\frac{45 s_{45}^6 s_{51}^2}{1024}-\frac{135 s_{45}^6 s_{51} s_{12}}{1024}-\frac{135 s_{45}^6 s_{12}^2}{1024}}{(D-3) s_{34}^4 (s_{23}-s_{45}-s_{51})^3}
+\frac{-\frac{9 s_{45}^6 s_{51}^2}{5120}-\frac{27 s_{45}^6 s_{51} s_{12}}{5120}-\frac{27 s_{45}^6 s_{12}^2}{5120}}{(D-1) s_{34}^4 (s_{23}-s_{45}-s_{51})^3}\\
+\frac{\frac{693 s_{45}^6 s_{51}^2}{5120}+\frac{2079 s_{45}^6 s_{51} s_{12}}{5120}+\frac{2079 s_{45}^6 s_{12}^2}{5120}}{(2 D-7) s_{34}^4 (s_{23}-s_{45}-s_{51})^3}
+\frac{-\frac{3 s_{45}^6 s_{51}^2}{1024}-\frac{9 s_{45}^6 s_{51} s_{12}}{1024}-\frac{9 s_{45}^6 s_{12}^2}{1024}}{s_{34}^4 (-s_{23}+s_{45}+s_{51})^3},
\end{multline}
which form a small part of our full reconstructed result.
Here $s_{ij}$ are the 5 kinematic variables of $R_*$.

The first feature is that 70\% of the 52,527 free parameters that we fitted turn out to be zero.
This can be understood by considering, for example, the first term in~\eqref{eq:some_reconstructed_terms}, and noting that a priori there was no reason for the numerator to only contain a term $\sim s_{45}^6 s_{12}^3$; in principle it could equally well have contained other mass-squared-dimension-9 combinations of $s_{45}$ and $s_{12}$, such as $s_{45}^2 s_{12}^7$.
For this reason, although expression~\eqref{eq:some_reconstructed_terms} contains only 16 numerator terms, to obtain them we actually fitted a total of 220 free parameters, of which 204 turned out to be zero.
Although at present our method does not exploit these additional simplifications, our strategy of reconstructing one partial-fractioned-term at is well-suited to exploiting them in the future.

Secondly, some of the numerators in our reconstructed result are linearly related to each other by a simple integer multiple.
For example, two of the numerators in the expression~\eqref{eq:some_reconstructed_terms}
\begin{equation}
n_1 = \frac{45 s_{45}^6 s_{51}^2}{1024}-\frac{135 s_{45}^6 s_{51} s_{12}}{1024}-\frac{135 s_{45}^6 s_{12}^2}{1024},
\end{equation}
\begin{equation}
n_2 = \frac{9 s_{45}^6 s_{51}^2}{5120}-\frac{27 s_{45}^6 s_{51} s_{12}}{5120}-\frac{27 s_{45}^6 s_{12}^2}{5120},
\end{equation}
can be observed to be equal up to an overall factor: $n_1 = 25 n_2$.
No such relations were exploited in obtaining the results in Table~\ref{tab:results}, but in future work if such relations can be observed prior to reconstruction, it would further reduce the number of free parameters and thus also reduce the number of probes required.

Thirdly, we notice that in some cases it is possible to combine several of our reconstructed terms and obtain a simpler expression. For example, if we combine together all the terms in expression~\eqref{eq:some_reconstructed_terms}, we obtain the following simple term:
\begin{equation}\label{eq:reconstructed_terms_combined}
-\frac{\frac{3}{512} D \left(D^2-4\right) s_{45}^6 (s_{51}+s_{12})^3}{(D-3) (D-1) (2 D-7) s_{34}^4 s_{51} (-s_{23}+s_{45}+s_{51})^3}.
\end{equation}
We should emphasise, however, that the first two features do not necessarily imply the third; indeed it was observed from examining other reconstructed terms that combining them in this way does not always simplify them.
The results in Table~\ref{tab:results} do not employ any such recombination of terms, and further study is required to understand which cases are amenable to such simplification and to devise a manner to exploit it during the reconstruction itself, rather than afterwards.



\section{Conclusion and outlook}\label{sec:conclusion}
In this conference proceedings paper, a method was presented for interpolation rational functions directly in partial-fractioned form.
This paper is based on our article Ref.~\cite{Chawdhry:2023yyx}, to which we refer the interested reader for more comprehensive explanations.
Our method employs $p$-adic numbers to perform the interpolation one partial-fractioned term at a time, harnessing the fact that the rational functions in loop calculations undergo a major simplification, often by orders of magnitude, under partial fractioning.
This simplification is not expected from a generic computer-algebraic viewpoint and so understanding its physical origin would be desirable.

Our technique was demonstrated by interpolating a complicated rational function at the edge of current calculational capabilities, namely the largest rational function in one of the largest IBP coefficients needed for non-planar 2-loop 5-point massless QCD amplitude calculations.
For this example it was shown that the number of numerical probes required using our technique is around 25 times smaller than in conventional $\F_p$-based techniques, and the final resulting expression is found to be $\mathcal{O}(100)$ times more compact.

We belive this provides a promising approach to calculating the rational functions in multi-loop calculations.
The techniques presented here may also assist in the search for analytical understanding of the physical origins of the simplification produced by partial fractioning such functions, an understanding which in turn could further inform the development of future interpolation strategies.
Finally, the notable even further simplifications that were observed post-hoc in section~\ref{sec:results} invite a deeper study from an analytic and physical viewpoint, and could ultimately also be incorporated into future interpolation techniques and strategies.

\section*{Acknowledgements}
We are grateful to Federico Buccioni, Fabrizio Caola, Fernando Febres Cordero, Stephen Jones, Stefano Laporta, Andrew M\textsuperscript{c}Clement, and Alex Mitov for helpful discussions. We also thank the authors of Ref.~\cite{Agarwal:2021vdh} for providing analytic expressions for the IBP coefficients used as examples in this work.
This work has been funded by the European Research Council (ERC) under the European Union's Horizon 2020 research and innovation programme (grant agreement no. 804394 \textsc{hipQCD}) and by the U.S. Department of Energy under grant DE-SC0010102.
We are also grateful to the Galileo Galilei Institute for hospitality and support during the scientific program on ``Theory Challenges in the Precision Era of the Large Hadron Collider'', where part of this work was performed.
%

\bibliographystyle{JHEP.bst}
\bibliography{ACAT_2024_proceedings_padic_reconstruction.bib}

\end{document}